\documentclass[preprint,12pt]{elsarticle}

\usepackage{graphicx}
\usepackage{dcolumn}
\usepackage{bm}
\usepackage{amssymb}

\begin{document}

\begin{frontmatter}

\title{Effects of the spike timing-dependent plasticity on the synchronisation 
in a random Hodgkin-Huxley neuronal network}

\author{R. R. Borges$^1$, F. S. Borges$^1$, A. M. Batista$^{1,2,*}$, 
E. L. Lameu$^1$, R. L. Viana$^3$, K. C. Iarosz$^4$, I. L. Caldas$^4$, M. A. F.
Sanju\'an$^5$}
\address{$^1$P\'os-Gradua\c c\~ao em Ci\^encias, Universidade Estadual de 
Ponta Grossa, 84030-900, Ponta Grossa, PR, Brazil.}
\address{$^2$Departamento de Matem\'atica e Estat\'istica, Universidade
Estadual de Ponta Grossa, 84030-900, Ponta Grossa, PR, Brazil.}
\address{$^3$Departamento de F\'isica, Universidade Federal do Paran\'a, 
81531-990, Curitiba, PR, Brazil.}
\address{$^4$Instituto de F\'isica, Universidade de S\~ao Paulo, 05315-970, 
S\~ao Paulo, SP, Brazil.}
\address{$^5$Departamento de F\'isica, Universidad Rey Juan Carlos, Tulip\'an 
s/n, 28933 M\'ostoles, Madrid, Spain.}
\cortext[cor]{Corresponding author: antoniomarcosbatista@gmail.com}

\date{\today}

\begin{abstract}
In this paper, we study the effects of spike timing-dependent plasticity on 
synchronisation in a network of Hodgkin-Huxley neurons. Neuron plasticity is a
flexible property of a neuron and its network to change temporarily or
permanently their biochemical, physiological, and morphological characteristics,
in order to adapt to the environment. Regarding the plasticity, we consider
Hebbian rules, specifically for spike timing-dependent plasticity (STDP), and 
with regard to network, we consider that the connections are randomly 
distributed. We analyse the synchronisation and desynchronisation according to 
an input level and probability of connections. Moreover, we verify that the
transition for synchronisation depends on the neuronal network architecture,
and the external perturbation level.
\end{abstract}

\begin{keyword}
plasticity \sep neuronal network \sep synchronisation
\end{keyword}

\end{frontmatter}


\section{Introduction}

The human brain contains about $10^{11}$ neurons \cite{chialvo2004}, and each 
neuron is connected to approximately $10^{4}$ other neurons \cite{gerstner}. 
These connections called synapses, are arranged in a highly complex network. 
They are responsible for neuronal communication and can be classified into two 
categories: electrical and chemical synapses \cite{bear2008, byrne2008}. In 
electrical synapses the transmission of information from one neuron to another 
is directly performed from the pre-synaptic cell to the post-synaptic cell via
gap junctions. In chemical synapses, the process occurs via neurotransmitters,
which cross the synaptic cleft and bind to receptors on the membrane of the
synaptic cell \cite{abbott}. Neurotransmitters may increase or decrease the 
probability of an action potential of a post-synaptic neuron, and the synapses 
are called excitatory or inhibitory, respectively \cite{purves2004}. 
Furthermore, the intensity of the chemical synapses can be modified, in other 
words, they can be minimised or potentiated \cite{dagostin2012}. The mechanism 
responsible for these adjustments is known as synaptic plasticity 
\cite{hebb1949}. 

The synaptic plasticity, that is, the ability of synapses to weaken or 
strenthen over time \cite{malenka2008} is an important property of the 
mammalian brain. In addition, the synaptic plasticity is also related to 
processes of learning and memory \cite{kelso1993,caporale2008}. This adjustment
of the intensities of the chemical synapses can be correlated with phenomena of
synchronisation of the neuronal firing \cite{rabinovich2003}. 

The occurrence of synchronisation in some specific areas of the brain may be 
associated with some diseases, such as the epilepsy and the Parkinson's disease 
\cite{abuhassan,modolo,hammond2007}. On the other hand, it is also responsible 
for some vital brain functions, such as processing of sensory information and 
motor function \cite{nini1995,uhlhaas2006}. 

Methods to suppress synchronisation have been proposed in neuroscience, as 
the introduction of external perturbations 
\cite{lameu2012,popovych2012,popovych2013}. Tass and collaborators have 
verified the possibility of desynchronisation in hippocampal neuronal 
populations through coordinated reset stimulation \cite{tass2009}. Meanwhile, 
Popovych and collaborators have found that the introduction of a perturbation 
in a globally connected neuronal network combined with synaptic plasticity can 
provide a positive contribution to the firing synchronisation 
\cite{popovych2013}. 

In this work, we study firing synchronisation in a random Hodgkin-Huxley 
neuronal network with plasticity according to spike timing-dependent plasticity
(STDP). This synaptic plasticity model adjusts the connection strengths by 
means of the temporal interval between pre-synaptic and post-synaptic spikes 
\cite{feldman2012,markram}. Bi and Poo have reported that the change in 
synaptic efficiency after several repetitions of the experiment is due to the 
time difference of firing \cite{bi1998,gerstner2010}. If one pre-synaptic spike 
precedes a post-synaptic spike, a long-term potentiation occurs, otherwise, a
long-term depression appears \cite{gray}.

A computational neuronal network specifies the connection architecture among 
neurons. A globally coupled Hodgkin-Huxley neuron model was also considered by 
Popovych and collaborators \cite{popovych2013}. They studied the 
synchronisation behaviour considering STDP, and found that the mean synaptic 
coupling presents a dependence on the input level. In this work, we consider a 
random neuronal network with STDP, and input, where the connections are 
associated with chemical synapses \cite{nordenfelt2013}. One main result is to 
show that spike synchonisation in a neuronal network, depending on the 
probability of connections, can be improved due to spike timing-dependent 
plasticity. This improvement is also observed when an external perturbation is 
applied on the network. Another important result is that the orientation of the 
connections among neurons with a different spike frequency affect the 
synchronised behaviour. 

This paper is organised as follows: in Section II we introduce the 
Hodgkin-Huxley neuronal model. In Section III, we show the random neuronal 
network. In Section IV, we study the synchronisation considering spike 
timing-dependent plasticity. Finally, in the last Section, we draw the 
conclusions.


\section{Hodgkin-Huxley neuronal network}

\subsection{Hodgkin-Huxley neuronal model}

\begin{figure}[htbp]
\begin{center}
\includegraphics[height=8cm,width=8cm]{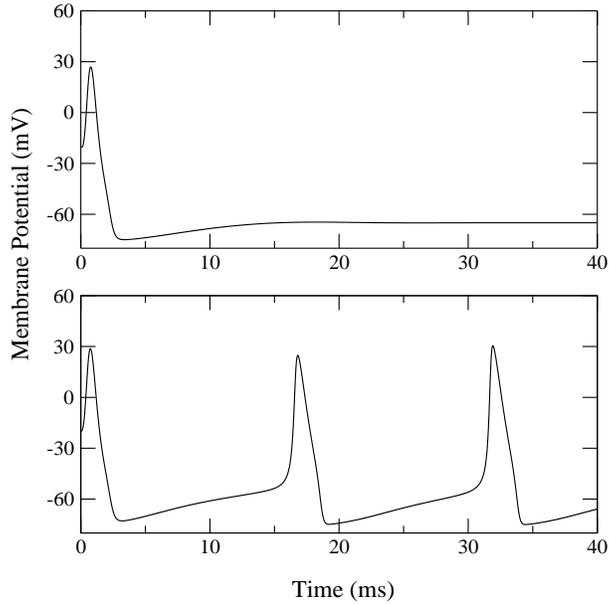}
\caption{Dynamic firing in the Hodgkin-Huxley model, where we consider
(a) $I=0.0\mu$A/cm$^2$ that shows a single spike with a subsequent resting 
state, and (b) $I=9.0\mu$A/cm$^2$ that presents a regime with periodic firing.}
\label{fig1}
\end{center}
\end{figure}

One of the most important models in computational neuroscience is the neuronal 
model proposed by Hodgkin and Huxley  \cite{hodgkin1952,izhikevich2004}. In 
this model, the mechanism of generation of an action potential was elucidated 
in a series of experiments with the squid giant axon. They found three
 different ions currents consisting of sodium (Na), potassium (K) and leak 
(L) mainly due to chlorine. Moreover, there are voltage-dependent channels
 for sodium, potassium that control the entry and exit of these ions through 
the cell. The model is composed of a system of four coupled differential 
equations given by
\begin{eqnarray}
\label{Equacoes_HH}
C\dot{V} & = & I-g_{\rm K}n^{4}(V-E_{\rm K})-g_{\rm Na}m^{3}h(V-E_{\rm Na}) \nonumber\\
& & -g_{\rm L}(V-E_{\rm L}),\\
\dot{n} & = & \alpha_{n}(V)(1-n)-\beta_{n}(V)n,\\
\dot{m} & = & \alpha_{m}(V)(1-m)-\beta_{m}(V)m,\\
\dot{h} & = & \alpha_{h}(V)(1-h)-\beta_{h}(V)h,
\end{eqnarray}
where $C$ is the membrane capacitance (measured in $\mu$F/cm$^2$), $V$ is the 
membrane potential (measured in mV), the function $m(V)$ and $n(V)$ are the 
variable of activation for sodium and potassium, and $h(V)$ is the variable of 
inactivation for sodium. The functions $\alpha_{n}$, $\beta_{n}$, $\alpha_{m}$, 
$\beta_{m}$, $\alpha_{h}$, $\beta_{n}$ are given by
\begin{eqnarray}
\alpha_{n}(V) & = & 0.01\frac{10-V}{\exp\left (\frac{10-V}{10}\right)-1},\\
\beta_{n}(V) & = & 0.125\exp\left(\frac{-V}{80}\right),\\
\alpha_{m}(V) & = & 0.1\frac{25-V}{\exp\left (\frac{25-V}{10}\right)-1},\\
\beta_{m}(V) & = & 4\exp\left(\frac{-V}{18}\right),\\
\alpha_{h}(V) & = & 0.07\exp\left(\frac{-V}{20}\right),\\
\beta_{n}(V) & = & \frac{1}{\exp \left(\frac{30-V}{10}\right)+1}.
\end{eqnarray}
The parameters $g$ and $E$ represent the conductance and reversal potentials 
for each ion. The constant $I$ is an external current density (measured in
$\mu$A/cm$^2$) that determines a regime of a single spike ($I=0.0\mu$A/cm$^2$), 
or a scheme with periodic spikes ($I=9.0\mu$A/cm$^2$), as illustrated in Fig. 
\ref{fig1}(a) and (b), respectively. Moreover, the spikes frequency increases 
when the constant $I$ increases. For instance, $I=9.0\mu$A/cm$^2$ and 
$I=10.0\mu$A/cm$^2$ approximately correspond to $67$Hz and $70$Hz, respectively.
The parameters that we use in this work are presented in Table 
\ref{parametros_HH} \cite{izhikevich2006}.
\begin{table}[htbp!]
\begin{center}
\caption{Parameters of the Hodgkin-Huxley neuronal model with a resting 
potential equal to $-65$mV.}
\begin{tabular}{|c|c|c|}
\hline 
Description                       & Parameter  & Values           \\ 
\hline 
Membrane capacity                 &   $C$      & 1 $\mu$F/cm$^{2}$ \\ 
\hline 
Reversal potential for Na         &   $E_{\rm Na}$  & 120 mV            \\
\hline 
Reversal potential for K          &   $E_{\rm K}$   & -12 mV           \\ 
\hline
Reversal potential for L          &   $E_{\rm L}$   &  10.6 mV         \\
\hline
Sodium conductance                &   $g_{\rm Na}$  &  120 mS/cm$^{2}$  \\
\hline
Potassium conductance             &   $g_{\rm K}$   &  36 mS/cm$^{2}$   \\
\hline
Leak conductance                  &   $g_{\rm L}$   &  0.3 mS/cm$^{2}$  \\ 
\hline
External current                  &   $I$      & 9.0 – 10.0 $\mu$A/cm$^{2}$  \\
\hline
\end{tabular} 
\label{parametros_HH}
\end{center}
\end{table}

\subsection{Network structure}

\begin{figure}[htbp]
\begin{center}
\includegraphics[height=6cm,width=6cm]{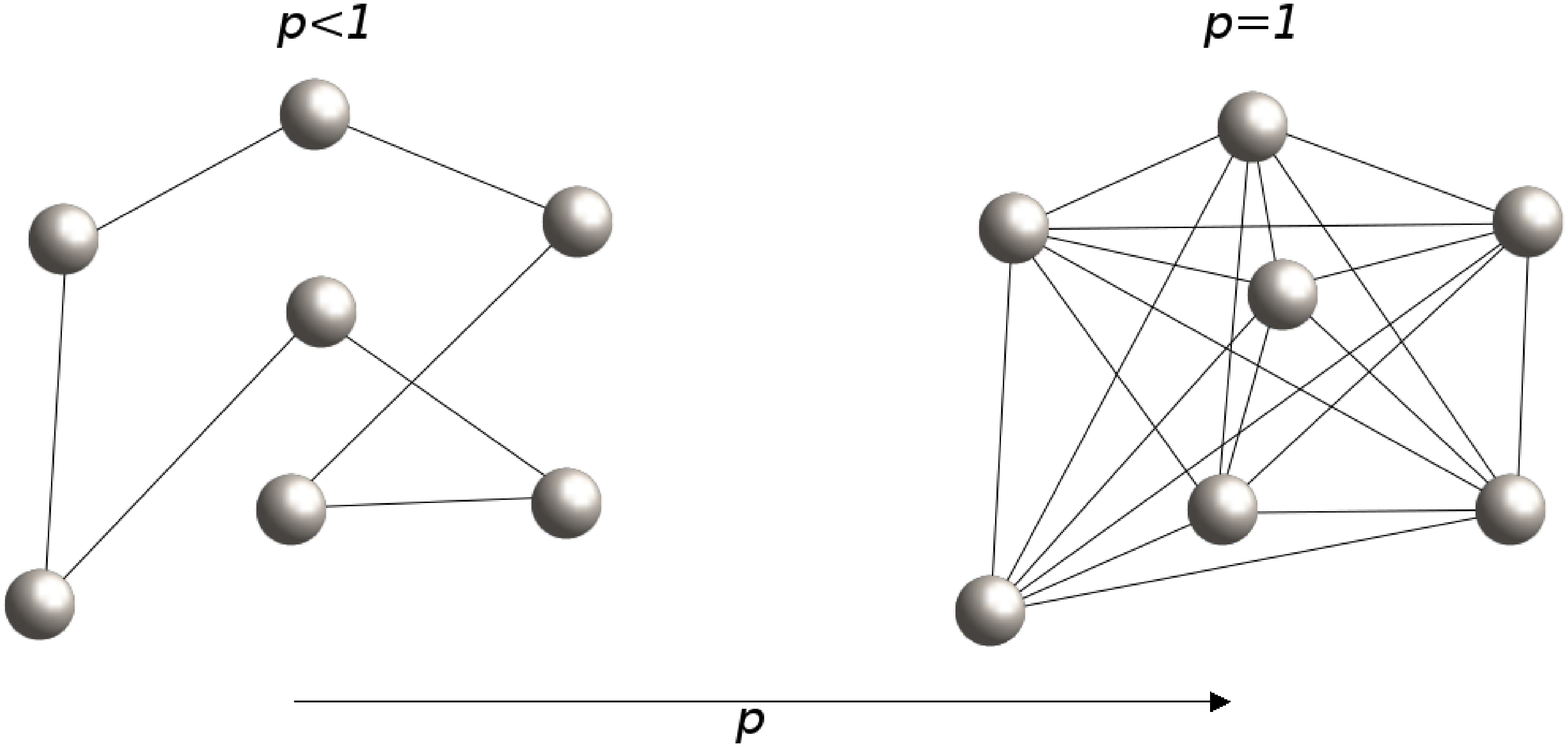}
\caption{Schematic representation of a network in which each neuron is 
connected to others by randomly chosen neurons with probability $p$.}
\label{fig2}
\end{center}
\end{figure}

Computational models of neuronal networks depend on the architecture, which 
specifies how neurons are connected and how the dynamics is applied to each 
unit or node. In this work, we consider a random network, that is, the network
is constructed by connecting neurons randomly 
\cite{erdos1959,nordenfelt,nordenfelt2014}. Each connection is included with 
probability independent from every other connection. Figure \ref{fig2} exhibits
a schematic representation of the neuronal network considered in this work. 
Each neuron is connected to others by randomly chosen neurons with probability 
$p$. When $p=1$ we have a global network, where all neurons are connected.

We consider a random neuronal network with chemical synapses where the 
connections are unidirectional, and the local dynamics is described by the
Hodgkin-Huxley model. The network is given by 
\begin{eqnarray}
C\dot{V_{\rm i}} & = & I_i-g_{\rm K}n^{4}(V_{\rm i}-E_{\rm K})-
g_{\rm Na}m^{3}h(V_{\rm i}-E_{\rm Na})- \nonumber \\
& & g_{L}(V_{i}-E_{L})+\frac{(V_{r}-V_{i})}{\omega}
\sum_{j=1}^{N}\varepsilon_{ij}s_{j}+\Gamma_i,
\end{eqnarray}
where $V_i$ is the membrane potential of neuron $i$ ($i=1,...,N$), $I_i$ is a
constant current density randomly distributed in the interval $[9.0,10.0]$, 
$\omega$ is the average degree connectivity, and $\varepsilon_{ij}$ is the 
coupling strength from the pre-synaptic neuron $j$ to the post-synaptic neuron 
$i$, that is, normally distributed with mean and variance equal to $0.1$ and 
$0.02$, respectively \cite{gray}. We consider an external perturbation 
$\Gamma_i$, so that each neuron receives an input with a constant intensity 
$\gamma$ during $1$ms. This input is applied with an average time interval 
around $14$ms. This value is approximately the inter-spike interval of a single
neuron. The neurons are excitatory coupled with a reversal potential 
$V_{r}=20$mV \cite{popovych2013}. The post-synaptic potential $s_{i}$ is given by
\cite{destexhe1994,golomb1993} 
\begin{equation}
\frac{ds_{i}}{dt}=\frac{5(1-s_{i})}{1+\exp\left(-\frac{V_{i}+3}{8}\right)}-s_{i}.
\end{equation}

\subsection{Spiking neurons synchronisation}

When identical neurons are coupled, the network may exhibit a complete 
synchronisation among spiking neurons, i.e., all neurons have identical time
evolution of their action potential. We do not consider identical neurons here,
and due to this fact, a complete synchronisation is not possible. However, a 
weak synchronisation may be observed.

As diagnostic of spikes synchronisation we use the order parameter given 
by \cite{kuramoto1975,kuramoto1984}
\begin{equation}
R=\left| \frac{1}{N}\sum_{j=1}^{N}\exp(i\psi_{j})\right| ,
\end{equation} 
where 
\begin{equation}
\psi_{j}(t)=2\pi m+2\pi\frac{t-t_{j,m}}{t_{j,m+1}-t_{j,m}},
\end{equation}
where $t_{j,m}$ denotes when a spike $m$ ($m=0,1,2,\dots$) of a neuron $j$ 
occurs ($t_{j,m}< t < t_{j,m+1}$). The beginning of each spike is considered
when $V_j>0$. If the spikes times are uncorrelated, their contribution to the 
result of the summation is small \cite{acebron}. However, in a globally 
synchronised state the order parameter magnitude asymptotes the unity. 

\begin{figure}[htbp]
\begin{center}
\includegraphics[height=9cm,width=8cm]{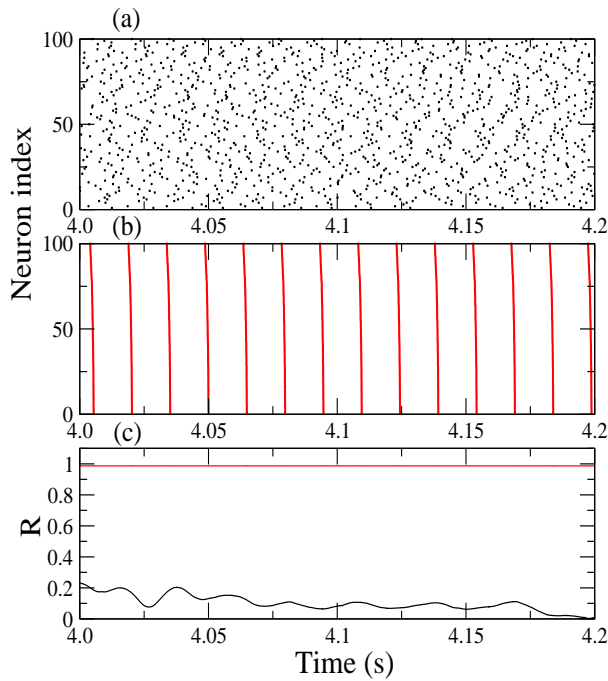}
\caption{(Colour online) Space-time plots of the membrane potentials for 
$N=100$, (a) $p=0.1$, and (b) $p=1.0$. In (c) the order parameter is calculated
for $p=0.1$ (black line), and $p=1.0$ (red line).}
\label{fig3}
\end{center}
\end{figure}

Figure \ref{fig3}(a) shows the spiking patterns for $p=0.1$, and without
external perturbation, that is, the neuronal network presents asynchronous 
dynamics, where the points correspond to spiking neurons, and the absent points
correspond to the resting neurons. For $p=1.0$, we have globally coupled 
neurons, and we can see that the network exhibits synchronised spiking, shown 
in Fig. 
\ref{fig3}(b). The time evolution of the order parameter is plotted in Fig. 
\ref{fig3}(c). When $p=0.1$ (black line), the network does not display 
synchronised behaviour, and as a result the order parameter is typically small 
with $R$ fluctuating around $0.1$. However, synchronised behaviour is observed 
for $p=1.0$ with order parameter values near unity (red line).

\begin{figure}[htbp]
\begin{center}
\includegraphics[height=6cm,width=7cm]{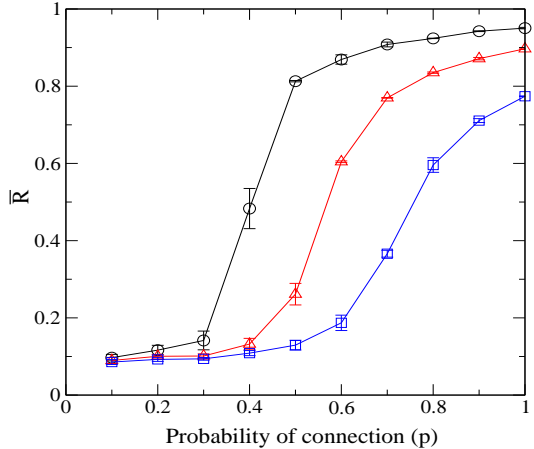}
\caption{(Colour online) Average order parameter as a function of the
probability, for $N=100$, for external perturbation with $\gamma=0$ (black 
circles), $\gamma=5$ (red triangles), and $\gamma=10$ (blue squares). The bars
represent the standard deviation.}
\label{fig4}
\end{center}
\end{figure}

We add an external perturbation ($\Gamma_i$) to analyse its effects on the
synchronous behaviour. This way we compute the time averaged magnitude of the 
order parameter, given by
\begin{equation}
\bar{R}=\frac{1}{t_{\rm fin}-t_{\rm ini}}\sum_{t=t_{\rm ini}}^{t_{\rm fin}} R(t),
\end{equation}
where the values of $\bar{R}$ have been computed by averaging over a temporal 
length of $10$s after discarding a transient of $490$s ($t_{\rm ini}=490$s and 
$t_{\rm fin}=500$s). In Fig. \ref{fig4} we can see the time averaged order 
parameter as a function of the probability for different input amplitudes 
($\gamma$). When the neuronal network has no input, it is possible to observe 
synchronised behaviour if the probability of connection is large enough (black 
circles). However, external inputs are able to desynchronise the spiking 
neurons, as shown in Fig. \ref{fig4} for $\gamma=5$ (red triangles) and 
$\gamma=10$ (blue squares), where the values of $\bar{R}$ are less than $0.9$.

\section{Spike timing-dependent plasticity}

One of the key principles of behavioural neuroscience is that experience can
modify the brain structure, that is known as neuroplasticity \cite{ramon1928}.
Although the idea that experience may modify the brain structure can probably 
be traced back to the 1890s \cite{bliis1973,bliss1993}, it was Hebb who made 
this a central feature in his neuropsychological theory \cite{hebb1961}.
 
\begin{figure}[htbp]
\begin{center}
\includegraphics[height=6cm,width=7cm]{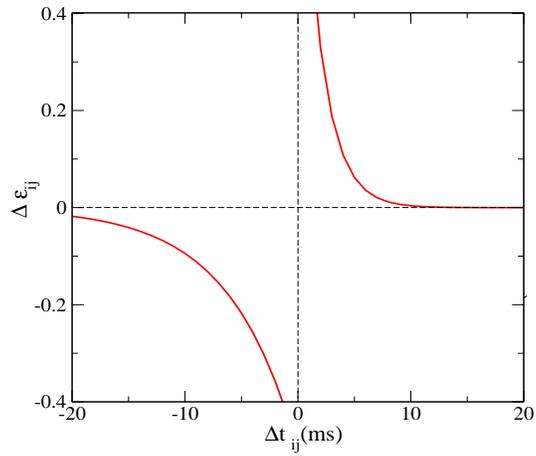}
\caption{Plasticity function (\ref{eqplast}) as a function of the difference of
spike timing of post- and pre-synaptic neuron.}
\label{fig5}
\end{center}
\end{figure}

With this in mind, we consider spike timing-dependent plasticity according to
the Hebbian rule. In this plasticity the coupling strength $\varepsilon_{ij}$
is adjusted based on the relative timing between the spikes of pre-synaptic
and post-synaptic neurons \cite{bi1998}.

\begin{equation}\label{eqplast}
\Delta \varepsilon_{ij}= \left\{
\begin{array}{ll}
\displaystyle A_{1}\exp(-\Delta t_{ij}/\tau_{1})\;,\;\Delta t_{ij}\geq 0 \\
\displaystyle -A_{2}\exp(\Delta t_{ij}/\tau_{2})\;,\;\Delta t_{ij} < 0 
\end{array}
\right. ,
\end{equation}
where $\Delta t_{ij}=t_{i}-t_{j}=t_{\rm pos}-t_{\rm pre}$. Figure \ref{fig5} exhibits
the result that is obtained from Eq. (\ref{eqplast}) for $A_{1}=1.0$, 
$A_{2}=0.5$, $\tau_{1}=1.8$ms, and $\tau_{2}=6.0$ms. 

\begin{figure}[htbp]
\begin{center}
\includegraphics[height=9cm,width=7cm]{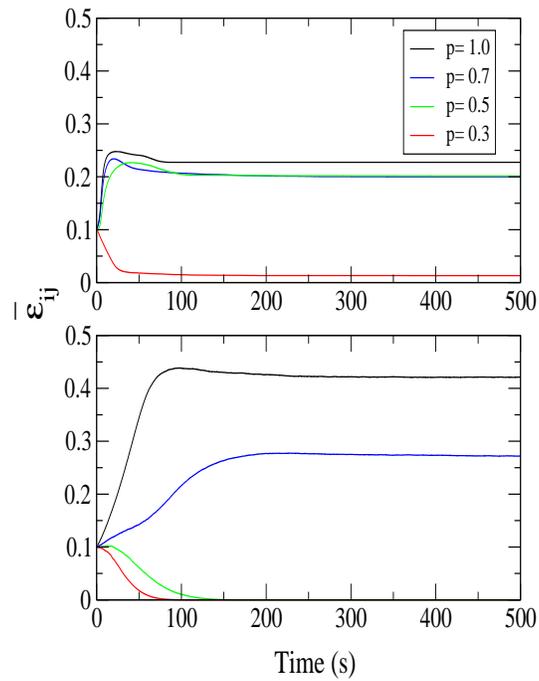}
\caption{(Colour online) Time evolution of the averaged coupling strength for 
$p=0.3$ (red line), $p=0.5$ (green line), $p=0.7$ (blue line), and $p=1.0$ 
(black line), where it is considered (a) $\gamma=0$, and (b) $\gamma=5$.}
\label{fig6}
\end{center}
\end{figure}

The initial synaptic weights $\varepsilon_{ij}$ are normally distributed with
mean and variance equal to 0.1 and 0.02, respectively. Then, they are updated 
according to Eq. (\ref{eqplast}), where $\varepsilon_{ij}\rightarrow
\varepsilon_{ij}+10^{-3}\Delta\varepsilon_{ij}$. In the absence of an external 
perturbation, we can verify by means of Fig. \ref{fig6}(a) that the averaged 
synaptic weights can be depressed ($p=0.3$) or potentiated ($p=0.5$, $p=0.7$, 
and $p=1.0$) depending on the probability of connections. If an input is 
applied on the neuronal network (Fig. \ref{fig6}b), the input can have a 
constructive effect on the synaptic weights ($p=0.7$ and $p=1.0$) or 
destructive effect ($p=0.3$ and $p=0.5$), depending on the probability of 
connections.

\begin{figure}[htbp]
\begin{center}
\includegraphics[height=7cm,width=8cm]{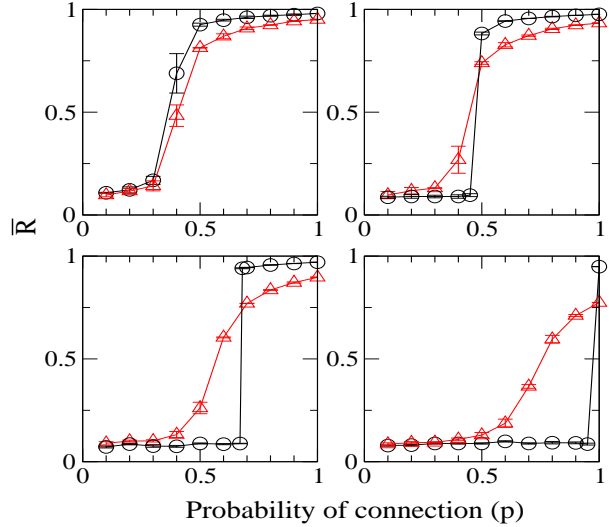}
\caption{(Colour online) Average order parameter versus probability of 
connection for the cases with (black circles) and without (red triangles)
STDP, where we consider (a) $\gamma=0$, (b) $\gamma=3$, (c) $\gamma=5$, and 
(d) $\gamma=10$. The bars represent the standard deviation.}
\label{fig7}
\end{center}
\end{figure}

The time averaged order parameter in terms of the probability of connection is 
showed in Fig. \ref{fig7}, considering with (black circles) and without (red
triangles) STDP. In the case for without external perturbation (Fig. 
\ref{fig7}a) we can see that the values of $\bar{R}$ without STDP are less than
with STDP, namely STDP is producing a positive effect on the synchronisation. 
Increasing the amplitude of the external perturbation, without spike 
timing-dependent plasticity, the desynchronisation is induced in the neuronal 
network (Fig. \ref{fig4}). However, considering STDP in the perturbed network 
is possible to observe alterations to the dynamic behaviour in relation to 
synchronised states. Figure \ref{fig7}(b) shows that the STDP enhances the
synchronisation for $p$ approximately greater than $0.5$ due to a constructive
effect on the dynamics of the synaptic weights. On the other hand, for $p$ less
than $0.5$ the STDP decreases the values of the time averaged order parameter,
as a result of depressed synaptic weights. Increasing the input intensity 
($\gamma=5$), it is possible to verify synchronisation when the network has 
neuroplasticity (Fig. \ref{fig7}c). We can also see an abrupt transition
from a desynchronised to a synchronised regime. When the input intensity 
increases, it is necessary to increase the probability of connection for the 
STDP counteracts the suppression of synchronisation. Figure \ref{fig7}(d) 
exhibits a situation such that the synchronisation is only obtained when the 
neuronal network presents a global coupling ($p=1$). With a strong input, the 
STDP does not lead the network to a potentiation of synaptic weights, and this 
way the synchronisation is suppressed by an external input.

\begin{figure}[htbp]
\begin{center}
\includegraphics[height=9cm,width=7cm]{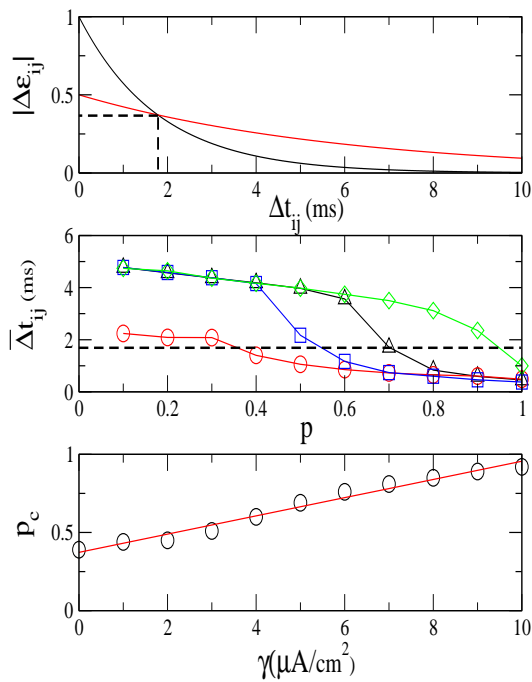}
\caption{(Colour online) (a) Absolute value of the plasticity function 
(\ref{eqplast}) as a function of the difference of spike timing of post- and 
pre-synaptic neuron, where the curves are the potentiation (red line) and the 
depression (black line). (b) Average time difference versus probability of 
connections for $\gamma=0$ (red circles), $\gamma=3$ (blue squares), 
$\gamma=5$ (black triangles), and $\gamma=10$ (green diamonds). (c) Critical 
probability $p_{c}$ as a function of the input level $\gamma$. The linear fit
 is given by the equation $p_{c}=0.06\gamma+0.37$.}
\label{fig8}
\end{center}
\end{figure}

The spike synchronisation depends on the probability of connections $p$ in a
way showing an abrupt transition. There is a critical point for $p=p_c$ that
can be found by means of the intersection between the curves of potentiation
and depression. Figure \ref{fig8}(a) exhibits the point of intersection with 
value of $\Delta t_{ij}$ approximately equal to $1.8$. For $\Delta t_{ij}>1.8$
the depression (red line) of the synaptic strength is larger than the 
potentiation (black line), while that for $\Delta t_{ij}<1.8$ the potentiation 
is larger than the depression. With this value of $\Delta t_{ij}$ we can obtain
$p_c$ plotting $\bar{\Delta t_{ij}}$ as a function of $p$, as shown in Fig. 
\ref{fig8}(b). It can be seen that the value of $\bar{\Delta t_{ij}}$ decays, 
and when cross the value $\bar{\Delta t_{ij}}\approx 1.8$ we have the critical 
value of the probability of connections. Moreover,  when $p$ increases not only 
$\bar{\Delta t_{ij}}$ decreases, but also the variance of the inter-spike 
intervals decreases. Then, we compute the critical probability as a function
of the input level (Fig. \ref{fig8}c), where we verify a linear increase 
given by the equation $p_{c}=0.06\gamma+0.37$.

\begin{figure}[htbp]
\begin{center}
\includegraphics[height=4cm,width=4cm]{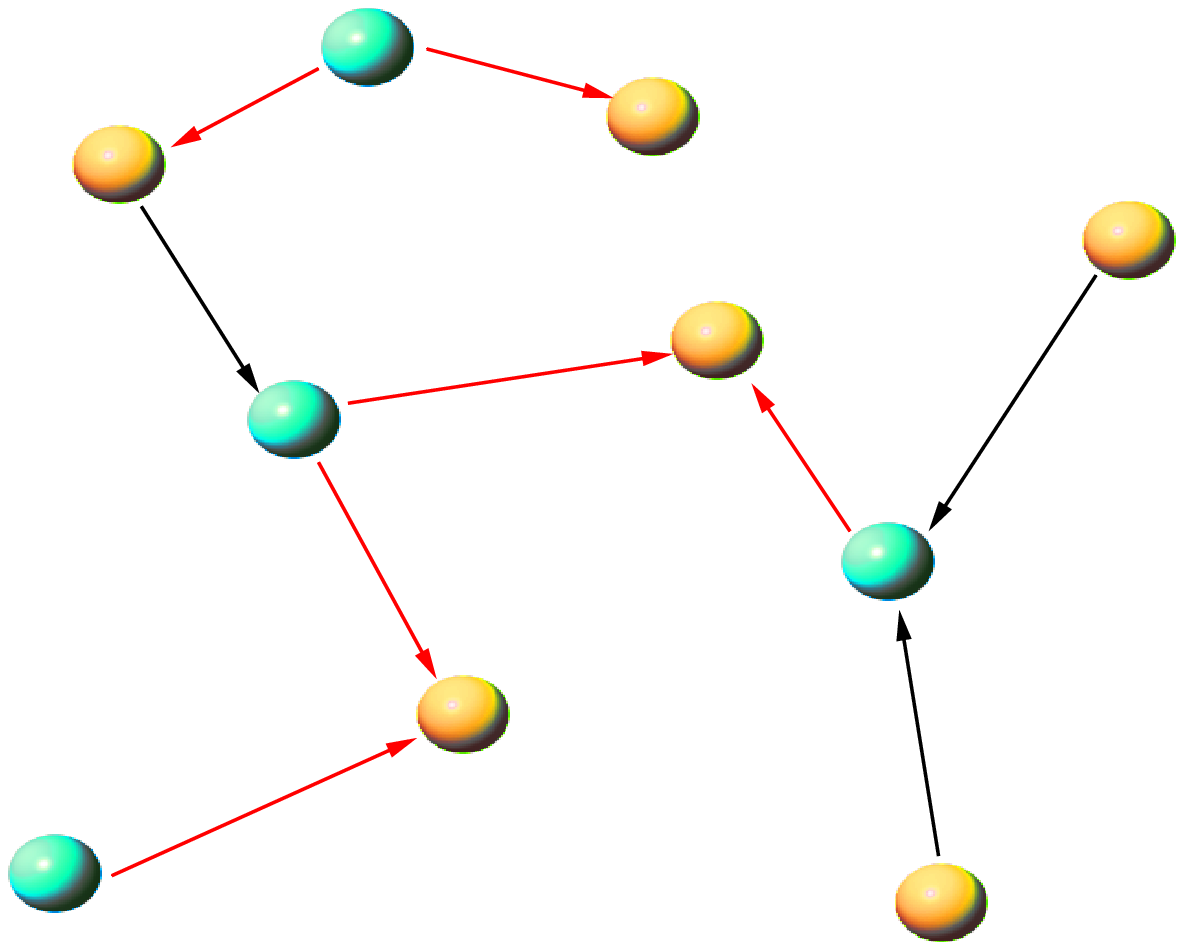}
\caption{(Colour online) Schematic representation of a network of neurons with 
high (cyan ball) and low (yellow ball) spiking frequency. The red arrows 
represent the connections between neurons from high to low frequency, while the
black arrows represent the connections from low to high frequency.}
\label{fig9}
\end{center}
\end{figure}

Figure \ref{fig7}(c) shows a discontinuous transition between the synchronised
and the desynchronised regime. Our aim is to understand how this discontinuous
transition appears when the probability of connections is varied. For this
reason, we build a network according to a schematic representation that is
showed in Fig. \ref{fig9}. The scheme represents a network of neurons with 
high (cyan ball) and low (yellow ball) spiking frequency. The red arrows 
represent the connections between neurons from high to low frequency, while the
black arrows represent the connections from low to high frequency.

\begin{figure}[htbp]
\begin{center}
\includegraphics[height=7cm,width=7cm]{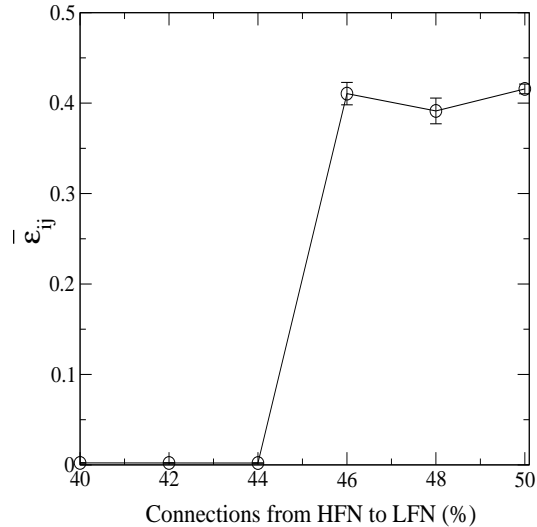}
\caption{Time averaged coupling strength versus percentage of connections from 
neurons with high frequency (HFN) to neurons with low frequency (LFN). We 
consider $p=0.47$ and $\gamma=3$.}
\label{fig10}
\end{center}
\end{figure}

Based on the schematic representation that is illustrated in Fig. \ref{fig9}, we
consider a neuronal network with  $N=100$, $p=0.47$, and $\gamma=3$. We 
separate the neuronal network into $50$ neurons with high frequency (values of 
$I_i$ within the range $[9.0,9.1]$) and $50$ neurons with low frequency (values
of $I_i$ within the range $[9.9,10]$). Figure \ref{fig10} exhibits the time 
averaged coupling strength as a function of the percentage of connections from 
neurons with high frequency to neurons with low frequency. We can see that
the time average coupling strength depends on the connections. Considering
the case for a small percentage of connections from HFN to LFN, the average 
coupling is small, indicating absence of synchronisation, a situation that 
changes with increasing the percentage of connections to a synchronised state.
This means that, when the coupling strengths increase, a desynchronised state 
can suddenly become synchronised. Consequently, the abrupt transition from
desynchronised to synchronised state, that is observed in Fig. \ref{fig7}, is
due to directed synapses among spiking neurons with high and low frequency.

\section{Conclusion}

We have been studying a neuronal network model with spiking neurons. We have 
chosen, as local dynamics, the Hodgkin-Huxley model due to the fact that it has 
essential features of spiking dynamics. The Hodgkin-Huxley model is a coupled 
set of nonlinear differential equations that describes the ionic basis of the 
action potential. These equations are able to reproduce biophysical properties
of the action potential.

We have used a random coupling architecture where the connections are randomly
distributed according to a probability. When the probability is equal to unity
we have a globally coupled network. The connections were considered 
unidirectional representing excitatory chemical synapses.

We have studied the effects of spike timing-dependent plasticity on the 
synchronisation in a Hodgkin-Huxley neuronal network. Studies about spike 
synchronisation are important to understand not only progressively degenerative
neurological disorders, but also processing of the sensory information. 
Popovych and collaborators \cite{popovych2013} showed that STDP combined with 
an external perturbation can improve the spike synchronisation in a globally 
neuronal network. The novelty in this paper is that we have considered a random
neuronal network and we have verified that the spike synchronisation depend on 
the probability of connections. Considering a strong external perturbation the 
spike synchronisation is suppressed. However, when there is STDP, depending on 
the probability of connections, the synchronisation in the perturbed network 
can be improved due to a constructive effect on the synaptic weights.

We have also shown that the direction of synapses has an important role on the 
effects of spike timing-dependent plasticity on the synchronisation in a random
Hodgkin-Huxley neuronal network. 


\section*{Acknowledgements}

This study was possible by partial financial support from the following 
Brazilian government agencies: CNPq, CAPES and FAPESP. Financial support by the
Spanish Ministry of Economy and Competitivity under project number 
FIS2013-40653-P is also acknowledged.


\end{document}